\documentclass[10pt,A4paper,conference]{IEEEtran}

\newtheorem{lemma}{\bf Lemma}
\newtheorem{theorem}{\bf Theorem}

\newtheorem{corollary}{\bf Corollary}

\newcommand{\cC}{\cal C}

\newcommand{\cF}{\cal F}
\newcommand{\cH}{\cal H}

\newcommand{\cR}{\cal R}

\newcommand{\cU}{\cal U}
\newcommand{\cX}{\cal X}

\def\squarebox#1{\hbox to #1{\hfill\vbox to #1{\vfill}}}

\begin{document}

\title{New Upper Bounds on $A(n,d)$}

\author{\authorblockN{Beniamin Mounits}
\authorblockA{Department of Mathematics\\
Technion-Israel Institute of Technology\\
Haifa 32000, Israel \\
Email: mounitsb@techunix.technion.ac.il} \and
\authorblockN{Tuvi Etzion}
\authorblockA{Department of Computer Science\\
Technion-Israel Institute of Technology\\
Haifa 32000, Israel \\
Email: etzion@cs.technion.ac.il} \and
\authorblockN{Simon Litsyn}
\authorblockA{Department of Electrical Engineering\\
Tel Aviv University\\
Ramat-Aviv 69978, Israel\\
Email: litsyn@eng.tau.ac.il}}
%

\maketitle

\begin{abstract}
Upper bounds on the maximum number of codewords in a binary code
of a given length and minimum Hamming distance are considered. New
bounds are derived by a combination of linear programming and
counting arguments. Some of these bounds improve on the best known
analytic bounds. Several new record bounds are obtained for codes
with small lengths.
\end{abstract}

\section{Introduction}
Let $A(n,d)$ denote the maximum number of codewords in a binary
code of length $n$ and minimum Hamming distance $d$. $A(n,d)$ is a
basic quantity in coding theory. Lower bounds on $A(n,d)$ are
obtained by constructions. For survey on the known lower bounds
the reader is referred to \cite{Lit}.

In this work we consider upper bounds on $A(n,d)$.
The most basic upper bound on $A(n,d)$, $d=2e+1$, is the sphere
packing bound, also known as the Hamming bound:
\begin{equation}\label{sphere-packing_bound}
A(n,2e+1) \leq \frac{2^n}{\sum_{i=0}^{e} {{n} \choose {i}}}~.
\end{equation}
Johnson \cite{Joh1} has improved the sphere packing bound. In his
theorem, Johnson used the quantity $A(n,d,w)$, which is the
maximum number of codewords in a binary code of length $n$,
constant weight $w$, and minimum distance $d$:
\begin{equation}
\label{john} A(n,2e+1) \leq \frac{2^n}{\sum_{i=0}^{e}{{n} \choose
{i}} + \frac{ {{n} \choose {e+1}} - {{2e+1} \choose {e+1}}
A(n,2e+2 ,2e+1)} {A(n,2e+2,e+1)}}~.
\end{equation}
In \cite{MoEtLi} a new bound was obtained:
\begin{equation}
\label{imprJohn} A(n,2e+1) \leq \frac{2^n}{\sum_{i=0}^{e}{{n}
\choose {i}} + \frac{ {{n+1} \choose {e+2}} - {{2e+2} \choose
{e+2}} A(n+1,2e+2,2e+2)} {A(n+1,2e+2,e+2)}} ~.
\end{equation}
This bound is at least as good as the Johnson bound for all values
of $n$ and $d$, and for each $d$ there are infinitely many values
of $n$ for which the new bound is better than the Johnson bound.

When someone is given specific, relatively small values, of $n$
and $d$, usually the best method to find upper bound on $A(n,d)$
is the linear programming (LP) bound. A summary about this method
and some new upper bounds appeared in \cite{MoEtLi}. However, the
computation of this bound is not tractable for large values of
$n$. In this work we will present new upper bounds on $A(n,2e+1)$,
$e\geq 1$.

Let ${\cF}_2 = \{0,1\}$ and let ${\cF}_2^n$ denote the set of all
binary words of length $n$. For $x,y \in {\cF}_2^n$, $d(x,y)$
denote the Hamming distance between $x$ and $y$. Given $x,y \in
{\cF}_2^n$ such that $d(x,y)=k$, we  denote by $p_{i,j}^{k}$ the
number of words $z\in{\cF}_2^n$ such that $d(x,z)=i$ and
$d(z,y)=j$. This number is independent of choice of $x$ and $y$
and equal to
\begin{displaymath}p_{i,j}^{k}=\left\{
\begin{array}{ll}
{k \choose \frac{i-j+k}{2}}{n-k \choose \frac{i+j-k}{2}}&
$if~$i+j-k$~is~even,~$~\\~\\
0& $if~$i+j-k$~is~odd.~$\\
\end{array} \right.
\end{displaymath}
If $x=y$, then $p_{i,j}^{0}=\delta_{i,j}v_{i}$, where $v_{i}={n
\choose i}$ is the number of words at distance $i$ from
$x\in{\cF}_2^n$, and $\delta_{i,j}=1$ if $i=j$ and zero otherwise.
We also denote $v=2^{n}$. The $p_{i,j}^{k}$'s are the intersection
numbers of the Hamming scheme and $v_{i}$ is the valency of the
relation $R_{i}$. For the connection between association schemes
and coding theory the reader is referred to \cite{Del1},
\cite[Chapter 21]{MaSl}.

An $(n,M,2e+1 )$ code $\cC$ is a nonempty subset of ${\cF}_2^n$ of
cardinality $M$ and minimum Hamming distance $2e+1$. For a word $x
\in {\cF}_2^n$, $d(x, {\cC} )$ is the Hamming distance between $x$
and $\cC$, i.e., $d(x, {\cC} ) = min_{c \in {\cC}} d(x,c)$. A word
$h \in {\cF}_2^n$ is called a $hole$ if $d(h,C)>e$ and
${\cH}=\{h\in {\cF}^{n}_{2}:~d(h,{\cC})>e\}$ is the set of all
holes. Clearly, we have
\begin{equation}\label{holescardin}
|{\cH}|=v-|{\cC}|V(n,e),
\end{equation}
where $V(n,e)=\sum_{j=0}^{e}v_{j}$ it is the volume of sphere of
radius $e$. The $distance~distribution$ of $\cC$ is defined as the
sequence $A_i = | \{ ( c_1 , c_2 ) \in {\cC} \times {\cC}:~ d( c_1
, c_2 ) =i \} | / | {\cC} |$ for $0 \leq i \leq n$ and $A_{i}(c)$
denote the number of codewords at distance $i$ from $c\in {\cC}$.
We also define the (non-normalized) holes distance distribution
$\{D_{i}\}_{i=0}^{n}$ by $D_{i}=|\{ (h_{1},h_{2}) \in {\cH} \times
{\cH}:~ d(h_{1},h_{2}) =i\}|$, and $D_{i}(h)$ denote the number of
holes at distance $i$ from $h\in {\cH}$. Finally, we define $NC(
h, {\cC}, \Delta )$ to be the number of codewords of $\cC$ at
distance $\Delta$ from a hole $h$.
\section{Holes Distance Distribution}
In the first theorem we state that for a given $(n,M,2e+1)$ code
$\cC$, the distance distribution of the holes is uniquely
determined by the distance distribution $\{A_{i}\}_{i=0}^{n}$ of the code $\cC$.~\\
\begin{theorem}
If ${\cC}$ is an $(n,M,2e+1)$ code with distance distribution
$\{A_{i}\}_{i=0}^{n}$, then
$$
D_{i}=vv_{i}+|{\cC}|\left({\cR}({\cC},i)-2V(n,e)v_{i}\right),
$$
for each $i$, $0\leq i\leq n$, where~\\
\begin{equation}
\label{Rvalues}
{\cR}({\cC},i)=\sum_{k=0}^{n}\left(\delta_{i,k}+2\sum_{j=1}^e
p_{i,j}^{k}+ \sum_{l=1}^{n}\sum_{m=1}^e p_{l,m}^{k}\sum_{j=1}^e
p_{i,j}^{l}\right)A_{k}.
\end{equation}
\end{theorem}~\\
\begin{corollary} \label{distdistrholes}
Let ${\cC}$ be an $(n,M,2e+1)$ code with distance distribution
$\{A_{i}\}_{i=0}^{n}$. If $\{q_{i}\}_{i=0}^{n}$ is a sequence of
real numbers, then
\begin{equation}\label{sum_D_i}
\sum_{i=0}^{n}q_{i}D_{i}=v\sum_{i=0}^{n}q_{i}v_{i}+|{\cC}|\sum_{i=0}^{n}q_{i}\left({\cR}({\cC},i)-2V(n,e)v_{i}\right).
\end{equation}
\end{corollary}~\\
By using Corollary \ref{distdistrholes}, for any given sequence
$\{q_{i}\}_{i=0}^{n}$  we obtain a linear combination of the
$D_{i}$'s. By finding a lower bound on this combination we can
obtain an upper bound on the size of ${\cC}$.~\\~\\
\underline{\emph{Example 1}}: Let $q_{0}=1$ and $q_{i}=0,~i>0$.
Clearly $v_{0}=1$ and by (\ref{Rvalues}) we have
${\cR}({\cC},0)=V(n,e)$. After substituting the trivial bound
$D_{0}\geq 0$ to (\ref{sum_D_i}) we obtain the sphere packing
bound (\ref{sphere-packing_bound}).

The sequence $\{q_{i}\}_{i=0}^{n}$ of Corollary
\ref{distdistrholes} will be called the \emph{holes distance
indices (HDI) sequence}. For convenience, in the rest of the paper
we will write $\{q_{i}\}$ instead of $\{q_{i}\}_{i=0}^{n}$. In the
next two sections we will find some good HDI sequences $\{q_{i}\}$
and develop methods to find lower bounds on
$\sum_{i=0}^{n}q_{i}D_{i}$.~\\
\section{HDI Sequences with Small Indices}\label{smallHDI}
In this section we consider HDI sequences, where nonzero $q_{i}$'s
correspond to small indices. The following lemma gives an
alternative expression for $D_{i}$.~\\
\begin{lemma}\label{D_i-expression} For each $i$, $0\leq i\leq n$,
$$
D_{i}=\sum_{h\in{\cH}}\left(v_{i}-\sum_{k=e+1}^{e+i}NC(h,{\cC},k)\sum_{j=0}^{e}p_{i,j}^{k}
\right).
$$
\end{lemma}~\\~\\
Given a sequence $\{q_{i}\}$, by using Lemma \ref{D_i-expression}
and (\ref{holescardin}) we estimate
$\sum_{i=0}^{n}q_{i}D_{i}$ in the following way.~\\
$$\sum_{i=0}^{n}q_{i}D_{i}=\sum_{i=0}^{n}q_{i}\sum_{h\in{\cH}}\left(v_{i}-\sum_{k=e+1}^{e+i}
NC(h,{\cC},k)\sum_{j=0}^{e}p_{i,j}^{k} \right)$$
$$
=\sum_{h\in{\cH}}\left(\sum_{i=0}^{n}q_{i}v_{i}-\sum_{i=0}^{n}q_{i}\sum_{k=e+1}^{e+i}
NC(h,{\cC},k)\sum_{j=0}^{e}p_{i,j}^{k}\right)
$$
\begin{equation}\label{new_estimate_D_i}
\geq
(v-|{\cC}|V(n,e))\left(\sum_{i=0}^{n}q_{i}v_{i}-{\xi}({\cC},\{q_{i}\})\right),
\end{equation}
where
\begin{equation}\label{psivalue}
{\xi}({\cC},\{q_{i}\})=\max_{h\in
{\cH}}\left\{\sum_{i=0}^{n}q_{i}\sum_{k=e+1}^{e+i}
NC(h,{\cC},k)\sum_{j=0}^{e}p_{i,j}^{k}\right\}.
\end{equation}
By combining (\ref{sum_D_i}) and (\ref{new_estimate_D_i}) we obtain~\\
\begin{theorem}\label{small-koef}
If ${\cC}$ is an $(n,M,2e+1)$ code with distance distribution
$\{A_{i}\}_{i=0}^{n}$, then
$$
|{\cC}|\leq
\frac{v}{V(n,e)+\frac{\sum_{i=0}^{n}q_{i}\left(V(n,e)v_{i}-{\cR}({\cC},i)\right)}{{\xi}({\cC},\{q_{i}\})}},
$$
provided ${\xi}({\cC},\{q_{i}\})$ is not zero, where
${\xi}({\cC},\{q_{i}\})$ is given by (\ref{psivalue}) and
${\cR}({\cC},i)$ is given by (\ref{Rvalues}).
\end{theorem}~\\
\underline{\emph{Example 2}}: Let $q_{1}=1$ and $q_{i}=0$ for
$i\neq 1$. From (\ref{Rvalues}) and (\ref{psivalue}) we have
$$
{\cR}({\cC},1)=V(n,e)v_{1}-p_{1,e}^{e+1}v_{e+1}+p_{1,e}^{e+1}p_{e+1,e}^{2e+1}A_{2e+1}
$$
and
$$
{\xi}({\cC},\{q_{i}\})=p_{1,e}^{e+1}\max_{h\in {\cH}}\left\{
NC(h,{\cC},e+1)\right\}
$$
Thus, using Theorem \ref{small-koef}, we obtain~\\
\begin{theorem}\label{dmui_Johnson}
If ${\cC}$ is an $(n,M,2e+1)$ code with distance distribution
$\{A_{i}\}_{i=0}^{n}$, then
\begin{equation}\label{dmui_Johnson_equation}
|{\cC}|\leq
\frac{v}{V(n,e)+\frac{v_{e+1}-p_{e+1,e}^{2e+1}A_{2e+1}}{\max_{h\in
{\cH}}\left\{ NC(h,{\cC},e+1)\right\}}}.
\end{equation}
\end{theorem}~\\~\\
By substituting $$A_{2e+1}\leq A(n,2e+2,2e+1)$$ and $$\max_{h\in
{\cH}}\left\{ NC(h,{\cC},e+1)\right\}\leq A(n,2e+2,e+1)$$ in
(\ref{dmui_Johnson_equation}) we obtain the Johnson upper bound
(\ref{john}).~\\~\\
\underline{\emph{Example 3}}: Let
$$q_{1}=\frac{p_{2,e}^{e+2}-p_{2,e-1}^{e+1}-p_{2,e}^{e+1}}{p_{1,e}^{e+1}}~,~q_{2}=1,$$ and
$q_{i}=0$ for $i \notin \{1,2\}$. From (\ref{Rvalues}) and
(\ref{psivalue}) we have
$$
q_{1}{\cR}({\cC},1)+q_{2}{\cR}({\cC},2)=V(n,e)(q_{1}v_{1}+q_{2}v_{2})-p_{2,e}^{e+2}(v_{e+1}
$$
$$
+v_{e+2}-(p_{e+1,e}^{2e+1}+p_{e+2,e-1}^{2e+1}+p_{e+2,e}^{2e+1})A_{2e+1}
-p_{e+2,e}^{2e+2}A_{2e+2})
$$
and
$$
{\xi}({\cC},\{q_{i}\})=p_{2,e}^{e+2}\max_{h\in {\cH}}\left\{
NC(h,{\cC},e+1)+NC(h,{\cC},e+2)\right\}
$$
Thus, using Theorem \ref{small-koef}, we obtain~\\
\begin{theorem}\label{dmui_impr_Johnson}
If ${\cC}$ is an $(n,M,2e+1)$ code with distance distribution
$\{A_{i}\}_{i=0}^{n}$, then
\begin{equation}\label{dmui_impr_Johnson_equation}
|{\cC}|\leq
\frac{v}{V(n,e)+\frac{v_{e+1}+v_{e+2}-\gamma}{\max_{h\in
{\cH}}\left\{ NC(h,{\cC},e+1)+NC(h,{\cC},e+2)\right\}}},
\end{equation}
where
$$
\gamma=(p_{e+1,e}^{2e+1}+p_{e+2,e-1}^{2e+1}+p_{e+2,e}^{2e+1})A_{2e+1}
+p_{e+2,e}^{2e+2}A_{2e+2}.
$$
\end{theorem}~\\~\\
By substituting $$A_{2e+1}+A_{2e+2}\leq A(n+1,2e+2,2e+2)$$ and
$$\max_{h\in {\cH}}\left\{ NC(h,{\cC},e+1)+NC(h,{\cC},e+2)\right\}$$ $$\leq A(n+1,2e+2,e+2)$$
in (\ref{dmui_impr_Johnson_equation}) we obtain the bound of
(\ref{imprJohn}).

Next, we want to improve the trivial bound on $A_{i}$ given by
$A_{i}\leq A(n,2e+2,i)$. We will find upper bounds on distance
distribution coefficients $A_{i}$'s using linear programming. For
an $(n,M,2e+1)$ code ${\cC}$ with distance distribution
$\{A_{i}\}_{i=0}^{n}$ let us denote by $LP[n,2e+1]$ the following
system of Delsarte`s linear constraints:
\begin{displaymath}
\left\{
\begin{array}{ll}
\sum_{i=0}^{n}A_{i}P_{k}(i)\geq 0 & \textrm {for $0\leq k\leq n$,}\\
0\leq A_{i}\leq A(n,2e+2,i) & \textrm {for $i=2e+1,2e+2,\ldots,n$,}\\
A_{0}=1, A_{i}=0 & \textrm {for $1\leq i< 2e+1$,}\\
\end{array} \right.
\end{displaymath}
where $P_{k}(i)=\sum_{j=0}^{k}(-1)^{j}{i \choose j}{{n-i} \choose
{k-j}}$ denote Krawtchouk polynomial of degree $k$. We also denote
$\tilde{n}=n+1$ and let $\{\tilde{A}_{i}\}_{i=0}^{\tilde{n}}$ be
the distance distribution of the $(n+1,M,2e+2)$ extended code
${\cC}_{e}$ which is obtained from the $(n,M,2e+1)$ code ${\cC}$
with distance distribution $\{A_{i}\}_{i=0}^{n}$ by adding an even
parity bit to each codeword of ${\cC}$. It's easy to verify that
for each $i$, $e+1\leq i\leq \lfloor \tilde{n}/2 \rfloor$,
\begin{equation}\label{coeff_dependence}
\tilde{A}_{2i}=A_{2i-1}+A_{2i}.
\end{equation}
For the even weight code ${\cC}_{e}$ of length $\tilde{n}$ and
distance $d=2e+2$ we denote by $LP_{e}[\tilde{n},2e+2]$ the
following system of Delsarte`s linear constraints:
\begin{displaymath}
\left\{
\begin{array}{ll}
\sum_{i=0}^{\tilde{n}}\tilde{A}_{i}P_{k}(i)\geq 0 & \textrm {for $0\leq k\leq \lfloor \tilde{n}/2 \rfloor$,}\\
0\leq \tilde{A}_{i}\leq A(\tilde{n},d,i) & \textrm {for $i=2e+2,2e+4,\ldots, 2\lfloor \tilde{n}/2 \rfloor$,}\\
\tilde{A}_{0}=1, \tilde{A}_{i}=0 & \textrm {for $1\leq i< 2e+2$.}\\
\end{array} \right.
\end{displaymath}
In some cases we will add more constraints to obtain some specific
bounds as in \cite{BBMOS,Hon,MoEtLi,vPu}.~\\
By Theorem \ref{dmui_impr_Johnson} we have that for an
$(n,M,2e+1)$ code ${\cC}$ with distance distribution
$\{A_{i}\}_{i=0}^{n}$ the following holds:
$$
|{\cC}|\leq \frac{2^n}{\sum_{i=0}^{e}{{n} \choose {i}}+\frac{
{{n+1} \choose {e+2}} - {{2e+2} \choose
{e+2}}(A_{2e+1}+A_{2e+2})}{A(n+1,2e+2,e+2)}}.
$$
Using (\ref{coeff_dependence}) we obtain~\\
\begin{theorem}\label{LP_impr_Johnson}
$$
A(n,2e+1)\leq \frac{2^n}{\sum_{i=0}^{e}{{n} \choose {i}}+\frac{
{{n+1} \choose {e+2}} - {{2e+2} \choose
{e+2}}max\{\tilde{A}_{2e+2}\}}{A(n+1,2e+2,e+2)}},
$$
where $max\{\tilde{A}_{2e+2}\}$ is taken subject to
$LP_{e}[\tilde{n},2e+2]$.
\end{theorem}~\\
For the next result we need the following theorem which is a
generalization of a theorem given by Best \cite{Be}.~\\
\begin{theorem}\label{gnrlzBest}
Let $\cC$ be a code of length $n$, minimum Hamming distance $d$,
and distance distribution $\{A_{i}\}_{i=0}^{n}$. Let
$\{p_{i}\}_{i=0}^{n}$ be a sequence of real numbers. Then there
exists a code $\cC'$ of length $n-1$, distance $d$ with distance
distribution $\{A_{i}'\}_{i=0}^{n-1}$ satisfying
\begin{equation}\label{gnrlzBest_equation}
\sum_{i=0}^{n}(n-i)p_{i}A_{i}\leq n\sum_{i=0}^{n-1}p_{i}A_{i}'.
\end{equation}
\end{theorem}~\\~\\
It was proved in \cite{RoVr} by using LP that for an even weight
code $\cC$ of length $\tilde{n}\equiv1(mod~4)$, distance $d=4$,
and distance distribution $\{\tilde{A}_{i}\}_{i=0}^{\tilde{n}}$,
\begin{equation}\label{Vroedt4}
\tilde{A}_{4}\leq
\frac{(\tilde{n}-1)(\tilde{n}-2)(\tilde{n}-3)}{24}.
\end{equation}
We substitute $p_{i}=\delta_{i,4}$ in (\ref{gnrlzBest_equation})
and (\ref{Vroedt4}) for the upper bound on $A_{4}'$ to obtain~\\
\begin{lemma}\label{maskanaVroedt4}
If $\cC$ is an even weight code of length
$\tilde{n}\equiv2(mod~4)$, distance $d=4$, and distance
distribution $\{\tilde{A}_{i}\}_{i=0}^{\tilde{n}}$, then
$$
\tilde{A}_{4}\leq \frac{\tilde{n}(\tilde{n}-2)(\tilde{n}-3)}{24}.
$$
\end{lemma}~\\
We take $e=1$, and $n\equiv9(mod~12)$ . Since
$A(n+1,4,3)=(n^2-3)/6$ for $n\equiv9(mod~12)$ \cite[p. 529]{MaSl},
it follows by Lemma \ref{maskanaVroedt4} and Theorem
\ref{LP_impr_Johnson} that~\\
\begin{theorem}
For $n\equiv9(mod~12)$
$$
A(n,3)\leq \frac{2^{n}}{n+3+\frac{4}{n^{2}-3}}.
$$
\end{theorem}~\\~\\
The previous best known bound $A(n,3)\leq 2^{n}/(n+3)$ for
$n\equiv1(mod~4)$ was obtained in \cite{BeBr} by LP. In
particular, we have $A(21,3)\leq 87348$ which improves on the
previous best known bound $A(21,3)\leq 87376$ \cite{MoEtLi}.~\\
\section{HDI Sequences with Large Indices}
We demonstrate another approach to estimating
$\sum_{i=0}^{n}q_{i}D_{i}$, where nonzero elements of $\{q_{i}\}$
correspond to large indices. For each $t$, $0\leq t\leq e$, we
denote
$$
E_{n-t}=\{h\in {\cH}|~NC(h,{\cC},n-t)=1\}.
$$
Note, that for any hole $h\in {\cH}$ we have $NC(h,{\cC},n-t)\in
\{0,1\}$,
where $0\leq t\leq e$.~\\
\begin{lemma}
For each $t$, $0\leq t\leq e$,
\begin{equation}\label{E_n-t_cardin}
|E_{n-t}|=|{\cC}|\left(v_{n-t}-\sum_{i=0}^{e+t}A_{n-i}\sum_{j=0}^{e}p_{n-t,j}^{n-i}\right).
\end{equation}
\end{lemma}~\\
Let $q_{n-1}=q_{n}=1$ and $q_{i}=0$ for $i \notin \{n-1,n\}$. If
$h\in E_{n-t}$ for $t\in \{0,1,\ldots,e-1\}$, then
\begin{equation}\label{h_in_E_n}
D_{n-1}(h)+D_{n}(h)=0.
\end{equation}
If $h\in E_{n-e}$, then
$$
D_{n-1}(h)+D_{n}(h)\geq n-e-(e+1)A(n-e,2e+2,e+1)
$$
\begin{equation}\label{h_in_E_n-e}
=n-e-(e+1)\lfloor \frac{n-e}{e+1}\rfloor.
\end{equation}
If for a given hole $h$ there exists no codeword at distance $k\in
\{n-e,n-(e-1),\ldots,n-1,n\}$, then
$$
D_{n-1}(h)+D_{n}(h)\geq n+1-(e+1)A(n,2e+2,e+1)
$$
\begin{equation}\label{h_not_in_E_n-e}
=n+1-(e+1)\lfloor \frac{n}{e+1}\rfloor.
\end{equation}
By combining (\ref{E_n-t_cardin})-(\ref{h_not_in_E_n-e}) with
Corollary \ref{distdistrholes} we obtain~\\

\begin{theorem}\label{bound_S=(n-1,n)}
$$
A(n,2e+1)\leq \frac{2^n}{2\sum_{i=0}^{e}{n \choose i}+\frac{{n
\choose e}\left((n-e)-(e+1)\lfloor
\frac{n-e}{e+1}\rfloor\right)-{\cU}(n)}{(e+1)\lfloor
\frac{n}{e+1}\rfloor}},
$$
where
$$
{\cU}(n)=\max\{{\cR}({\cC},n-1)+{\cR}({\cC},n)
$$
$$
-\left(n+1-(e+1)\lfloor \frac{n}{e+1}\rfloor
\right)\sum_{t=0}^{e-1}\sum_{i=0}^{e+t}A_{n-i}\sum_{j=0}^{e}p_{n-t,j}^{n-i}
$$
$$
-(e+1)\left(1+\lfloor \frac{n-e}{e+1}\rfloor-\lfloor
\frac{n}{e+1}\rfloor\right)\sum_{i=0}^{2e}A_{n-i}\sum_{j=0}^{e}p_{n-e,j}^{n-i}\},
$$
subject to $LP[n,2e+1]$ and ${\cR}({\cC},n-1)$, ${\cR}({\cC},n)$
are given by (\ref{Rvalues}).
\end{theorem}~\\
By Theorem \ref{bound_S=(n-1,n)} we obtain $A(22,3)\leq 172361$
and $A(24,5)\leq 47538$ which improve the previous best known
bounds $A(22,3)\leq 173015$ \cite{MoEtLi} and $A(24,5)\leq 48008$
\cite{Sch}.

Let $n$ be even integer and let
$e=1$. By Theorem \ref{bound_S=(n-1,n)} and (\ref{coeff_dependence}) we obtain~\\
\begin{theorem}\label{bound_S=(n-1,n)_e=1}
If $n$ is an even integer, then
$$
A(n,3)\leq \frac{2^n}{2n+3-\frac{{\cU}(n)}{n}},
$$
where
$$
{\cU}(n)=\max\{6\tilde{A}_{\tilde{n}-3}+3n\tilde{A}_{\tilde{n}-1}\}
$$
subject to $LP_{e}[\tilde{n},4]$.
\end{theorem}~\\
Using LP we can prove the following lemma.~\\
\begin{lemma}\label{newbound_n=10(12)}
If ${\cC}$ is an even weight code of length $\tilde{n}\equiv
11(mod~12)$, distance $d=4$, and distance distribution
$\{\tilde{A}_{i}\}_{i=0}^{\tilde{n}}$, then
$$
6\tilde{A}_{\tilde{n}-3}+3(\tilde{n}-1)\tilde{A}_{\tilde{n}-1}\leq
\frac{(\tilde{n}-1)(\tilde{n}-2)(\tilde{n}+4)}{\tilde{n}+2}.
$$
\end{lemma}~\\
Therefore, by Theorem \ref{bound_S=(n-1,n)_e=1} and Lemma
\ref{newbound_n=10(12)} we have~\\
\begin{theorem}
For $n\equiv 10(mod~12)$
$$
A(n,3)\leq \frac{2^n}{n+2+\frac{8}{n+3}}.
$$
\end{theorem}~\\~\\
The previous best known analytic bound
$$
A(n,3)\leq \frac{2^{n}}{n+2+\frac{2(n+14)}{n^2+n-8}}
$$
was obtained by (\ref{imprJohn}).

By similar arguments, if $\{q_{i}\}$ is a sequence with $q_{i}=0$,
except for $q_{n-2}=q_{n-1}=q_{n}=1$, we obtain the
following bound.~\\
\begin{theorem}\label{bound_S=(n-2,n-1,n)}
If ${\cC}$ is an $(n,M,2e+1)$ code, then
$$
|{\cC}|\leq \frac{2^n}{2\sum_{i=0}^{e}{n \choose
i}+\phi-\frac{{\cU}(n)}{{e+2 \choose 2} A(n+1,2e+2,e+2)}},
$$
where
$$
\phi=\frac{{n+1 \choose e}\left({n+1-e \choose 2}-{e+2 \choose 2}
A(n+1-e,2e+2,e+2)\right)}{{e+2 \choose 2} A(n+1,2e+2,e+2)}
$$
and
$$
{\cU}(n)=\max\{\sum_{i=n-2}^{n}{\cR}({\cC},i)-(1+{n+1 \choose 2}
$$
$$
-{e+2 \choose 2}A(n+1,2e+2,e+2)
)\sum_{t=0}^{e-2}\sum_{i=0}^{e+t}A_{n-i}\sum_{j=0}^{e}p_{n-t,j}^{n-i}
$$
$$
-(1+e(n-e)+{e+1 \choose 2}-{e+2 \choose 2}(A(n+1,2e+2,e+2)
$$
$$
-A(n+1-e,2e+2,e+2)))\sum_{t=e-1}^{e}\sum_{i=0}^{e+t}A_{n-i}\sum_{j=0}^{e}p_{n-t,j}^{n-i}\},
$$
subject to $LP[n,2e+1]$, and ${\cR}({\cC},n-2)$,
${\cR}({\cC},n-1)$, ${\cR}({\cC},n)$ are given by (\ref{Rvalues}).
\end{theorem}~\\
Applying Theorem \ref{bound_S=(n-2,n-1,n)} we obtain $A(21,3)\leq
87333$ which is better than the best previously known bound (see
Section \ref{smallHDI}).~\\
\section{Generalization for Arbitrary Metric Association Schemes}
We can generalize our approach to arbitrary metric association
scheme $({\cX},{\cR})$ with distance function $d$, which consists
of a finite set ${\cX}$ together with a set ${\cR}$ of $n+1$
relations defined on ${\cX}$ with certain properties. For the
complete definition and brief introduction to the association
schemes, the reader is referred to \cite[Chapter 21]{MaSl}. We
extend the definitions from the first section as follows.
$|{\cX}|=v$ is the number of points of a finite set ${\cX}$,
$v_{i}$ is the valency of the relation $R_{i}$, and
$p_{i,j}^{k}$'s are the intersection numbers of the scheme. A code
${\cC}$ is a nonempty subset of ${\cX}$ with minimum distance
$2e+1$. The definitions related to holes and distance distribution
are easily generalized. The results of (\ref{holescardin}), Lemma
\ref{D_i-expression}, Theorems \ref{D_i-expression} through
\ref{dmui_impr_Johnson}, and Corollary \ref{distdistrholes} are
valid for arbitrary metric association schemes.

As an example we consider the Johnson scheme. In this scheme
${\cX}$ is the set of all binary vectors of length $n$ and weight
$w$. Note, that in this scheme the number of relations is $w+1$
and $n$ has different meaning. The distance between two vectors is
defined to be the half of the Hamming distance between them. One
can verify, that $v={n \choose w}$, $v_{i}={w \choose i}{n-w
\choose i}$, and $p_{i,j}^{k}$ is given by
$$
\sum_{l=0}^{w-k}{w-k \choose l}{k \choose w-i-l}{k \choose
w-j-l}{n-w-k \choose i+j+l-w}.
$$
Denote by $T(w_{1},n_{1},w_{2},n_{2},d)$ the maximum number of
binary vectors of length $n_{1}+n_{2}$, having mutual Hamming
distance of at least $d$, where each vector has exactly $w_{1}$
ones in the first $n_{1}$ coordinates and exactly $w_{2}$ ones in
the last $n_{2}$ coordinates. By substituting $$\max_{h\in
{\cH}}\left\{ NC(h,{\cC},e+1)\right\}\leq T(e+1,w,e+1,n-w,4e+2)$$
in (\ref{dmui_Johnson_equation}), we obtain the following bound.~\\
\begin{theorem}\label{Johnson-Johnson}
$$
A(n,4e+2,w)\leq \frac{{n \choose w}}{\sum_{i=0}^{e}{w \choose
i}{n-w \choose i} +\frac{{w \choose e+1}{n-w \choose e+1}-{2e+1
\choose e}^{2}{\cU}_{w}(n)}{T(e+1,w,e+1,n-w,4e+2)}},
$$
where
$$
{\cU}_{w}(n)=max\{A_{2e+1}\},
$$
subject to Delsarte`s linear constraints for Johnson scheme (see
\cite[Theorem 12, p. 666]{MaSl} ).
\end{theorem}~\\
Applying Theorem \ref{Johnson-Johnson} for $e=1$ we obtain the
following improvements (the values in the parentheses are the best
bounds previously known \cite{AVZ1}, \cite{Sch}): $A(19,6,7)\leq
519~(520)$, $A(22,6,11)\leq 5033~(5064)$, $A(26,6,11)\leq
42017~(42080)$.

We would like to remark, that LP can be applied for upper bounds
that obtained by centering a spheres around a
codewords. We give an example of such bound.~\\
\begin{theorem}\label{Centering_spheres} $A(n,10,w)\leq$
$$
\frac{{n \choose w}}{\sum_{i=0}^{2}{w \choose i}{n-w \choose
i}+\frac{{w \choose 3}{n-w \choose 3}}{T(3,w,3,n-w,10)}+\frac{{w
\choose 4}{n-w \choose 4}}{T(4,w,4,n-w,10)}-{\cU}_{w}(n)},
$$
where
$$
{\cU}_{w}(n)=max\{\frac{225}{T(4,w,4,n-w,10)}A_{2e+2}
$$
$$
+\left(\frac{100}{T(3,w,3,n-w,10)}+\frac{50n-475}{T(4,w,4,n-w,10)}\right)A_{2e+1}\},
$$
subject to Delsarte`s linear constraints for Johnson scheme.
\end{theorem}~\\
By Theorem \ref{Centering_spheres} we have: $A(23,10,9)\leq
78~(81)$, $A(24,10,9)\leq 116~(119)$, $A(25,10,9)\leq 157~(158)$,
$A(27,10,9)\leq 293~(299)$, $A(28,10,10)\leq 785~(821)$.

\section*{Acknowledgment}
The work of Beniamin Mounits was supported in part by grant no.
263/04 of the Israeli Science Foundation.

The work of Tuvi Etzion was supported in part by grant no. 263/04
of the Israeli Science Foundation.

The work of Simon Litsyn was supported in part by grant no. 533/03
of the Israeli Science Foundation.

\end{document}